# The Rouse Ring Chain with Attractive Harmonic Potential of Spherical Symmetry.

Nail Fatkullin[1], Carlos Mattea[1], Kevin Lindt[1], Siegfried Stapf[1] and Margarita Kruteva[2]

[1]*Technische Physik II/Polymerphysik, Technische Universität Ilmenau, D-98684 Ilmenau, Germany*

[2]*Jülich Centre for Neutron Science (JCNS-1) and Institute for Biological Information Processing (IBI-8), Forschungszentrum Jülich GmbH, 52428 Jülich, Germany*

*Corresponding authors: nfatkull@gmail.com*

**Abstract**

The static and dynamic properties of a cyclic Rouse chain modified by the introduction of an effective, spherically symmetric, attracting potential of entropic nature is studied. It is shown that a relatively weak potential can lead to a strong contraction of the polymer chain: the radius of gyration becomes much smaller compared to the size of free cyclic chain. The pronounced decrease in the terminal relaxation time of cyclic macromolecules compared to the Rouse relaxation time leads to a lengthening of the time interval for the transition to the normal, i.e. the Fickian, diffusion regime, generating a pseudo-plateau at increasing molecular mass for the time dependence of segmental mean squared segmental displacement.

## 1. Introduction.

In recent years, interest in polymeric systems containing ring macromolecules has increased markedly (see, for example, [1-16] and the literature cited therein). As shown e.g. by rheology and neutron scattering experiments, the absence of end segments in cyclic macromolecules leads to serious differences in their conformations and dynamics in comparison with linear analogues of equal molecular mass. It also has an essential impact on the biology: the DNA in the genome forms cyclic structures that are much more compact than their linear analogues. In particular, in melts linear macromolecules have conformations that are, with sufficient accuracy, close to ideal Gaussian conformation, i.e. without intra-chain excluded volume interactions, [17-22] and the mean square of their radius of gyration depends linearly on the molecular weight $R_{g,lin}^{2}=\frac{Nb^{2}}{6}$, where $b$ is the length of Kuhn segment and $N$ is the number of Kuhn segments per macromolecule. In the case of melts of cyclic macromolecules, the absence of end segments generates entropic interactions preventing mutual penetration of macromolecules into each other, which leads to a significant weakening of the molecular-mass dependence of the mean-square radius of gyration $R_{g,ring}^2 \propto b^2 N^{\nu}$ with $\nu \approx \frac{2}{3}$ as has been shown by simulations [6-7]. In other words, these additional topological constraints appear to be similar to effective intramolecular attractive interactions related to the center of mass of the ring, leading to more compact, globular conformations. The latter circumstance makes the analytical description of the structure and dynamics of cyclic macromolecules more complicated compared to their linear analogs, since even at the level of intramolecular interactions the non-Gaussian distribution of conformations gives rise to nonlinear equations of motion.

In the theory of dynamic properties of linear polymers, the Rouse model plays an important supporting role because it is exactly solvable [17-22]. Although it does not include entanglement effects, it can be at least phenomenologically modified, for example, by constraints for chain motions only inside the tube for times smaller than the terminal relaxation time, as in the reptation tube model, to apply it satisfactorily even for highly entangled systems.

For the case of ring polymer melts, the Rouse model can be applied directly using appropriate boundary conditions for the ends of linear polymer chains, see for example [16]. In this case, ring macromolecules have a conformational distribution corresponding to a freely connected ring whose characteristic size is only $2^{1/2}$ times smaller than the size of linear chains containing the same number of Kuhn segments and the mean squared radius of gyration scales as $R_{g,ring}^2 = \frac{1}{12} b^2 N$. In other words, this simplified approach does not take into account the effects of contraction of polymer rings in melts associated with the above-mentioned topological constraints on mutual penetrations.

This contraction effect can be considered by introducing a harmonic potential of spherical symmetry which attracts any of the ring segments to the ring centre of mass with the strength constant as a parameter. After this modification, the Rouse model modified by a spherical attracting harmonic potential is still exactly analytically tractable and can play the role of a reference model for the dynamical properties of polymer rings. As far as we know, this model, although amenable to exact solution, has not yet been described in the literature. However, a similar model containing a modification of the Rouse model with an axially symmetric harmonic two-dimensional potential to study various aspects of the reptation model has been considered in [23,24]. In this manuscript we present the Rouse model modified by a three-dimensional harmonic potential to describe the dynamics of a ring polymer, which in this context is of independent scientific interest.

## 2. Theoretical part.

Consider a polymer ring formed of N>>1 Kuhn segments, then the effective Hamiltonian, more precisely the effective intramolecular potential energy, is defined by the following expression:

$$H = \frac{3kT}{2}\sum_n \left( \frac{1}{b^2}\left(\frac{\partial \vec{r}_n}{\partial n}\right)^2 + \frac{\left(\vec{r}_n - \vec{r}_{cm}\right)^2}{\tilde{R}^2} \right), \qquad (1)$$

where $\tilde{R}$ is a parameter of the potential. The first term on the right-hand side of relation (1) describes the traditional effectively intramolecular entropic interactions of an ideal polymer chain (see, for example, [17-22]), while the second term, which has the structure of a harmonic potential, is related to the effects of entropic contraction of the polymer chain caused by topological constraints of mutual penetration of cyclic macromolecules into each other. Note also that because N>>1 the continuous limit and summation over n is equivalent to integration over n. The parameter of the potential $\tilde{R}$ is defined in a way, that in the case of an isolated segment, its mean square distance from the minimum of potential is equal to

$$\left\langle r^2 \right\rangle = \tilde{R}^2 \quad . \qquad (2)$$

In the following we will call $\tilde{R}$ the characteristic radius of the harmonic potential. The larger this quantity, the weaker the potential.

The equation of motion for a Rouse ring is as follows:

$$\zeta \frac{\partial \vec{r}_n(t)}{\partial n} = \frac{3kT}{b^2}\frac{\partial^2}{\partial n^2}\vec{r}_n(t) - \frac{3k_BT}{\tilde{R}^2}\left(\vec{r}_n(t) - \vec{r}_{cm}(t)\right) + \vec{f}_n^{\,L}(t), \qquad (3)$$

where $\zeta$ is the segmental friction coefficient, $\vec{f}_n^{\,L}(t)$ is the stochastic Langevin force acting on segment with number *n* at time moment *t*, $\vec{r}_{cm}(t) = \frac{1}{N}\sum_n \vec{r}_n(t)$ is the position vector of the center of mass of the considered polymer ring at time moment *t*, $\vec{r}_n(t)$ is the position vector of the segment with number *n* at time moment *t*, $k_BT$ is the Boltzmann constant multiplied by the absolute temperature.

The ring architecture of the polymer chain is taken into account by means of the following boundary conditions:

$$\vec{r}_0(t) \equiv \vec{r}_N(t). \qquad (4)$$

These boundary conditions lead to the following Fourier decomposition:

$$\vec{r}_n(t) = \vec{X}_0(t) + \sum_{p=1,,,N/2} \left( 2\vec{X}_p(t)\cos\left(\frac{2\pi}{N}pn\right) + 2\vec{Y}_p(t)\sin\left(\frac{2\pi}{N}pn\right) \right). \tag{5}$$

The normal relaxation Rouse modes of the ring macromolecule can be calculated from expression (5) by the inverse Fourier transform:

$$\begin{aligned} \vec{X}_p(t) &= \frac{1}{N}\int_0^N \vec{r}_n(t)\cos\left(\frac{2\pi}{N}pn\right)dn \\ \vec{Y}_p(t) &= \frac{1}{N}\int_0^N \vec{r}_n(t)\sin\left(\frac{2\pi}{N}pn\right)dn \end{aligned} . \tag{6}$$

By substituting the decomposition (6) into relation (1), we express the Hamiltonian through the normal Rouse modes of the ring macromolecule:

$$H = \frac{3kT}{2}\sum_{p=1,...N/2}\left\{\frac{8\pi^2}{Nb^2}p^2 + \frac{2N}{\tilde{R}^2}\right\}\left(\vec{X}_p^2(t) + \vec{Y}_p^2(t)\right). \tag{7}$$

Assuming the different normal modes fluctuate independently of each other and have a normal Gaussian distribution, we obtain the following relation for their mean-square values:

$$\frac{1}{\left\langle \vec{X}_p^2 \right\rangle_{eq}} = \frac{1}{\left\langle \vec{Y}_p^2 \right\rangle_{eq}} = \left\{\frac{8\pi^2 p^2}{Nb^2} + \frac{2N}{\tilde{R}^2}\right\}. \tag{8}$$

Further, from relations (5), (8) and considering orthogonality of different normal modes, the following expression for the mean squared radius of gyration of the ring macromolecule is obtained:

$$\begin{aligned} R_g^2 &\equiv \frac{1}{N}\int_0^N \left\langle \left(\vec{r}_n(t)\right) - \vec{X}_0(t)^2 \right\rangle \\ &= 2\sum_p \left\{ \left\langle \vec{X}_p^2 \right\rangle_{eq} + \left\langle \vec{Y}_p^2 \right\rangle_{eq} \right\} = 2\sum_p \left\{ \frac{1}{\frac{8\pi^2 p^2}{Nb^2} + \frac{2N}{\tilde{R}^2}} + \frac{1}{\frac{8\pi^2 p^2}{Nb^2} + \frac{2N}{\tilde{R}^2}} \right\}. \\ &= 2\sum_{p=1...N/2} \frac{1}{\frac{4\pi^2 p^2}{Nb^2} + \frac{N}{\tilde{R}^2}} \end{aligned} \tag{9}$$

For a very weak harmonic potential $\frac{N^2b^2}{4\pi^2} \ll \tilde{R}^2$ the polymer ring has an ideal, not perturbed size (e.g. see [17]):

$$R_{g,ideal}^2 = \frac{Nb^2}{12}. \qquad (10)$$

For large Rouse rings one observes a globular state:

$$\begin{aligned}
R_g^2 &= 2\sum_{p=1\ldots N/2} \frac{1}{\frac{4\pi^2 p^2}{Nb^2} + \frac{N}{\tilde{R}^2}} \approx 2\int_1^{N/2} \frac{dp}{\frac{4\pi^2 p^2}{Nb^2} + \frac{N}{\tilde{R}^2}} \\
&= \frac{\tilde{R}b}{\pi}\left( arctg\left(\frac{\pi\tilde{R}}{b}\right) - arctg\left(\frac{2\pi\tilde{R}}{Nb}\right)\right) , \\
&= \begin{cases} \frac{1}{2}\tilde{R}b \quad if \quad \frac{Nb}{2\pi} \gg \tilde{R} \gg \frac{b}{\pi} \\ \\ \tilde{R}^2 \quad if \quad \tilde{R} \ll \frac{b}{\pi} \end{cases}
\end{aligned} \qquad (11)$$

where the approximation $arctg\left(x\right) = \begin{cases} x & if \quad x \ll 1 \\ \\ \pi/2 & if \quad x >> 1 \end{cases}$ for an infinitely large ring N>>1 was used.

The limit $\tilde{R} \ll \frac{b}{\pi}$ has mainly academic interest. For the interesting case $\frac{Nb}{2\pi} \gg \tilde{R} \gg \frac{b}{\pi}$ one finds that $R_g^2 = \frac{1}{2}\tilde{R}b \ll \tilde{R}^2$. Recall that $\tilde{R}$ is the inverse measure of the strength of the harmonic potential and equal to the typical distance at which a potentially isolated polymer segment will be separated from the center, see expression (2). The equation $R_g = \sqrt{\frac{\tilde{R}b}{2}} \ll \tilde{R}$ indicates that the ring size is much smaller due to the linear connectivity of joined segments. This effect is directly related to the linear coupled structure of the polymer ring and reflects the fact that there are neighboring segments in close proximity to any segment, resulting in an enhanced attraction by the center of potential. It is relevant to point out here that despite the constraint $\frac{Nb}{2\pi} \gg \tilde{R}$, the harmonic potential may in fact remain weak because N>>1.

Figure 1 illustrates the noted dependence of the square of the radius of gyration of the macromolecule on the potential parameter $\tilde{R}$ and the number of Kuhn segments $N$.

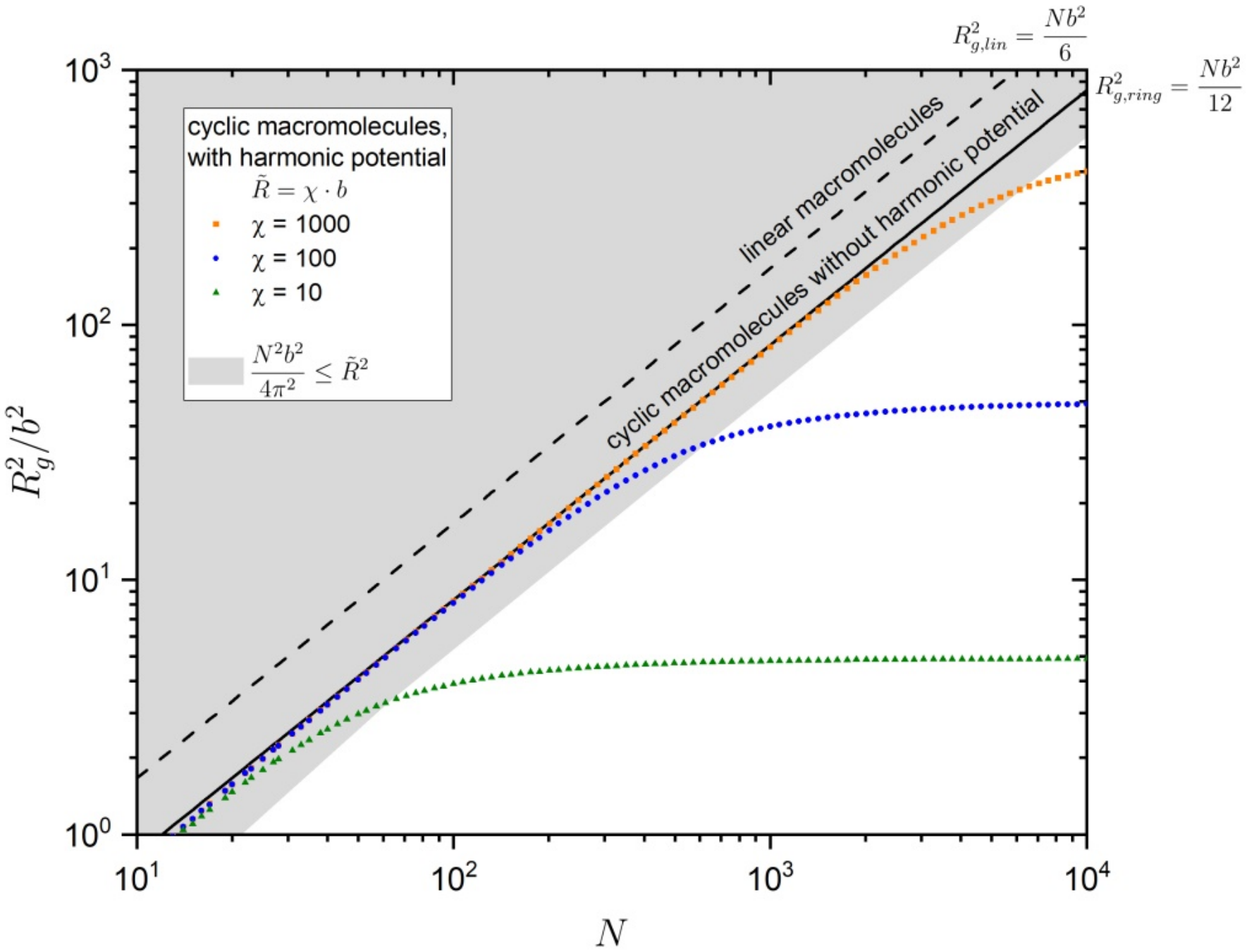

Fig. 1. Graphical illustration of the relations (9)-(11). The harmonic potential parameter  is considered as an independent variable of the number of Kuhn segments in the ring macromolecule .

Expressions for the decay rates of the normal modes, i.e., the inverse relaxation times, can be obtained by differentiating relation (5) in time using the equations of motion (3). The result is as follows:

$$\frac{1}{\tau_p} = \frac{4}{\tau_s}\left\{\left(\frac{p}{N}\right)^2 + \frac{1}{4\pi^2}\frac{b^2}{\tilde{R}^2}\right\} = \frac{1}{2\pi^2\tau_s N}\frac{b^2}{\left\langle \vec{X}_p^2 \right\rangle} = \frac{1}{2\pi^2\tau_s N}\frac{b^2}{\left\langle \vec{Y}_p^2 \right\rangle}, \tag{12}$$

with the segmental relaxation time $\tau_s = \dfrac{\zeta b^2}{3\pi^2 kT}$.

Note that the relaxation times of the long-wave Rouse modes with $p \ll \dfrac{1}{2\pi}\dfrac{bN}{\tilde{R}} = \dfrac{1}{4\pi}\dfrac{b^2 N}{\left\langle R_g^2 \right\rangle_{eq}}$ cease to depend on the mode number. Instead, they depend mainly on the parameter of the harmonic potential.

The mean squared displacement of ring segments can be derived from expression (5):

$$\left\langle \left( \vec{r}_n(t) - \vec{r}_n(0) \right)^2 \right\rangle = \left\langle \left( \vec{X}_0(t) - \vec{X}_0(0) \right)^2 \right\rangle + 4 \sum_{p=1,\dots N/2} \left[ \left\langle \vec{X}_p^2 \right\rangle + \left\langle \vec{Y}_p^2 \right\rangle \right] \left( 1 - \exp\left\{ -\frac{t}{\tau_p} \right\} \right). \quad (13)$$

The first contribution on the right-hand part of expression (13) relates to the mean squared displacement of the center of mass, whereas the second part, i.e. the sum over the normal modes $p = 1, \dots N/2$, describes motion in the coordinate system connected with this center of mass.

For times $\tau_s \ll t \ll \overline{\tau}_1 \equiv \dfrac{4\pi^2 \tau_s R_g^4}{b^4}$ the second term in the right-hand part of (13) dominates and anomalous Rouse diffusion is observed:

$$\begin{aligned} \left\langle \left( \vec{r}_n(t) - \vec{r}_n(0) \right)^2 \right\rangle &\approx 4 \sum_{p=1,\dots N/2} \left[ \left\langle \vec{X}_p^2 \right\rangle + \left\langle \vec{Y}_p^2 \right\rangle \right] \left( 1 - \exp\left\{ -\frac{t}{\tau_p} \right\} \right) \\ &= 8 \sum_{p=1,\dots N/2} \left\langle \vec{X}_p^2 \right\rangle \left( 1 - \exp\left\{ -\frac{t}{\tau_p} \right\} \right) \approx \frac{2b^2}{\pi^{3/2}} \left( \frac{t}{\tau_s} \right)^{1/2} \end{aligned} . \quad (14)$$

For longer times $\overline{\tau}_1 = \dfrac{4\pi^2 \tau_s R_g^4}{b^4} \ll t \ll \overline{\tau}_1^* = \pi^2 \tau_s \dfrac{R_g^2 N}{b^2}$ the same term still dominates, but the time dependence disappears and a plateau is reached:

$$\left\langle \left( \vec{r}_n(t) - \vec{r}_n(0) \right)^2 \right\rangle \approx 4 \sum_{p=1,\dots N/2} \left[ \left\langle \vec{X}_p^2 \right\rangle + \left\langle \vec{Y}_p^2 \right\rangle \right] = 2R_g^2. \quad (15)$$

Only for longer times $t \gg \pi^2 \tau_s \dfrac{R_g^2 N}{b^2}$ the first term on the right-hand part of expression (13), describing normal diffusion of the ring's center of mass, starts to dominate:

$$\left\langle \left( \vec{r}_n(t) - \vec{r}_n(0) \right)^2 \right\rangle \approx \left\langle \left( \vec{X}_0(t) - \vec{X}_0(0) \right)^2 \right\rangle = 6 \frac{b^2}{3\pi^2} \frac{t}{\tau_s N}. \quad (16)$$

In Figure 2, a graphical illustration of the formulas (13)-(16) is presented.

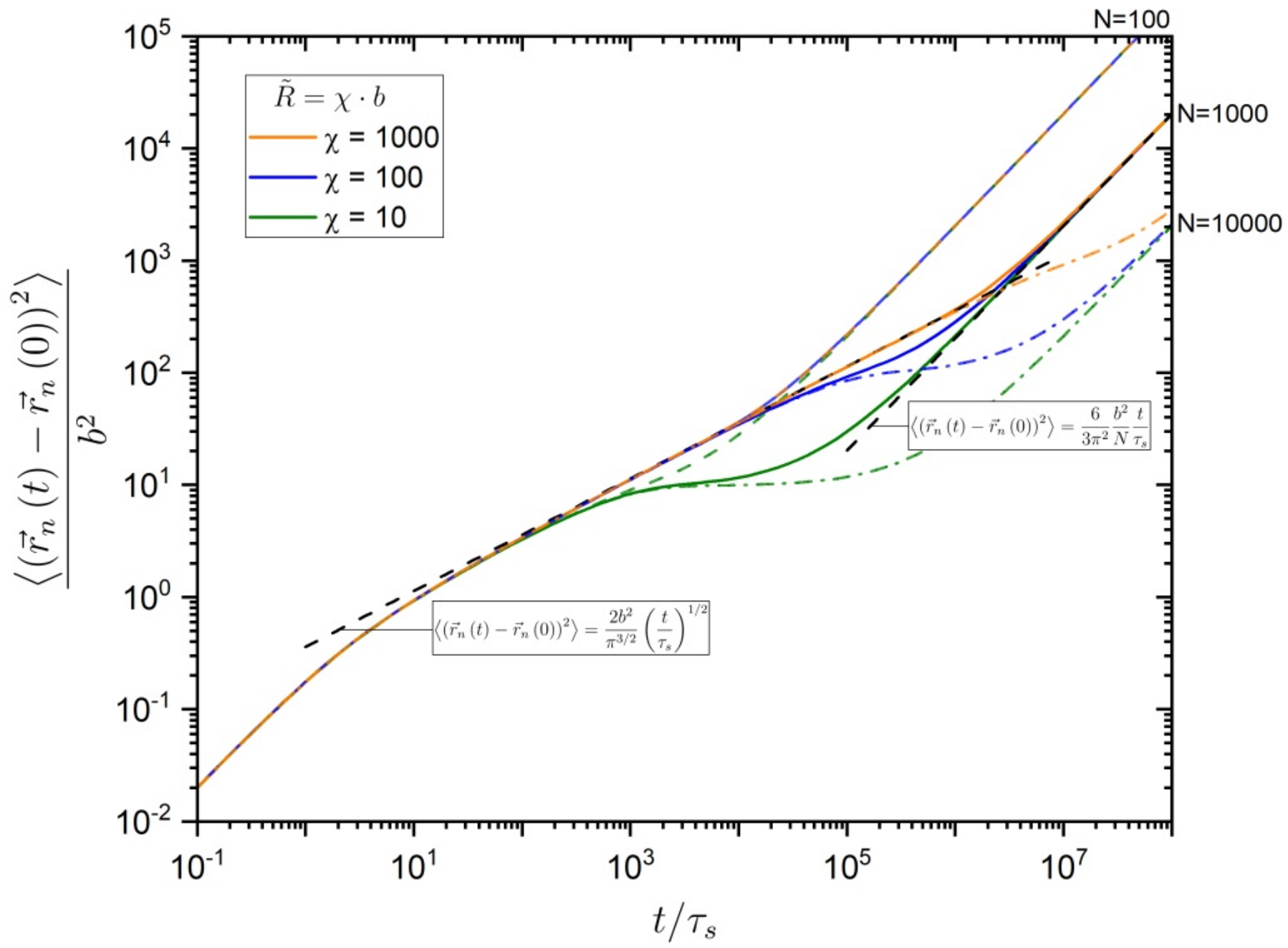

Fig.2. Graphical illustration of the relations (13)-(16). The harmonic potential parameter $\tilde{R}$ is again considered as an independent variable of the number of Kuhn segments in the ring macromolecule.

The incoherent dynamical structure factor of ring macromolecules, which is relevant for experiments such as neutron scattering, is given by the following relation:

$$S^{incoh}(k;t) \equiv \frac{1}{N} \sum_n \left\langle \exp\left\{ i\vec{k}\left(\vec{r}_n(t) - \vec{r}_n(0)\right)\right\} \right\rangle = \frac{1}{N} \sum_n \left\langle \exp\left\{ -\frac{1}{6} k^2 \left\langle \left(\vec{r}_n(t) - \vec{r}_n(0)\right)^2 \right\rangle \right\} \right\rangle . \tag{17}$$

while the coherent dynamical structure factor is

$$\begin{aligned} S^{coh}(k;t) &\equiv \frac{1}{N} \sum_{n,m} \left\langle \exp\left\{ i\vec{k}\left(\vec{r}_n(t) - \vec{r}_m(0)\right)\right\} \right\rangle = \frac{1}{N} \sum_{n,m} \left\langle \exp\left\{ -\frac{1}{6} k^2 \left\langle \left(\vec{r}_n(t) - \vec{r}_m(0)\right)^2 \right\rangle \right\} \right\rangle \\ &= \frac{1}{N} \sum_{n,m} \varphi_{nm}(k,t) \end{aligned} , \tag{18}$$

with

$$\varphi_{nm}(k,t) \equiv \left\langle \exp\left\{ i\vec{k}\left(\vec{r}_n(t) - \vec{r}_m(0)\right)\right\}\right\rangle$$

$$= \exp\left\{ -\frac{k^2\left\langle \left(\vec{X}_0(t) - \vec{X}_0(0)\right)^2 \right\rangle + 8\sum_{p=1,..N/2}\left\langle \vec{X}_p^2 \right\rangle \left(1 - \exp\left\{-\frac{t}{\tau_p}\right\}\cos\left(\frac{2\pi p}{N}(n-m)\right)\right)}{6} \right\} \quad . \qquad (19)$$

## 3. Discussion.

We investigated the properties of a simple model, which allows us to carry out a sufficiently complete and accurate analytic study: the Rouse ring chain modified by adding an attractive harmonic potential of spherical symmetry. Despite significant simplifications, the model has important nontrivial properties.

The model clearly illustrates the high sensitivity of the polymer ring conformation to the intensity of the attractive potential $U\{\vec{r}_n\} = \frac{3k_BT}{2}\sum_n \frac{\left(\vec{r}_n - \vec{r}_{cm}\right)^2}{\tilde{R}^2}$ for polymer rings containing a large number of Kuhn segments N>>1. As it follows from relation (11), at $\frac{Nb}{2\pi} \gg \tilde{R}$ the polymer rings experience strong contraction in comparison with the ideal state, and their characteristic linear dimensions become much smaller than the ideal rings. Indeed, when this condition is satisfied, the linear dimensions of the polymer ring are smaller than the analogous linear dimensions in the ideal ring state: $R_g = \frac{1}{2}\tilde{R}b \ll \sqrt{\frac{b^2N}{12}} \equiv R_{g.ideal}$.

It is instructive to discuss the density distribution of the segments belonging to the ring itself, at position $\vec{\tilde{r}}$ with respect to its center of mass:

$$\tilde{\rho}_s\left(\vec{\tilde{r}}\right) \equiv \sum_n \left\langle \delta\left(\vec{\tilde{r}} - \vec{r}_n + \vec{r}_{cm}\right)\right\rangle. \qquad (20)$$

Since in a ring macromolecule, all segments are equivalent to each other and in our simplified model the distribution of normal Rouse modes is assumed to be Gaussian, this value is equal to:

$$\tilde{\rho}_s\left(\vec{\tilde{r}}\right)=\frac{N}{\left(\frac{2\pi}{3}R_g^2\right)^{3/2}}\exp\left(-\frac{3}{2}\frac{\tilde{r}^2}{R_g^2}\right). \qquad (21)$$

In particular, at the very center of mass of the ring the concentration of own segments reaches the maximum value:

$$\tilde{\rho}_s\left(\vec{\tilde{r}}=0\right)=\frac{N}{\left(\frac{2\pi}{3}R_g^2\right)^{3/2}}. \qquad (22)$$

If a polymeric macromolecule containing $N$ Kuhn segments is contracted to a spherical globule with the radius $\tilde{R}_{sf}$ such that this globule contains only its own segments, their concentration will be equal to the concentration of polymeric segments in the melt $\rho_s$:

$$\rho_s=\frac{N}{\frac{4\pi}{3}\tilde{R}_{sf}^3}. \qquad (23)$$

By dividing ratio (22) by ratio (23), we obtain a value indicating what fraction of the globule's own segments density relative to the density of all segments at the globule's center of mass:

$$\frac{\tilde{\rho}_s(0)}{\rho_s}=\sqrt{\frac{6}{\pi}}\left(\frac{\tilde{R}_{sf}}{R_g}\right)^3. \qquad (24)$$

Let us consider experimental data for ring macromolecules of PEO with molecular mass M=96000 Da [25]: $\tilde{R}_{sf}=32\overset{0}{A}$, $R_g=49\overset{0}{A}$ (from SANS experiments), then $\frac{\tilde{\rho}_s(0)}{\rho_s}\approx 0.385$. This result is numerically very close to recent computer simulations [14] and the expression (21) at least qualitatively coincides with Fig. 3b from that work.

In this very simplified approach the parameter $\tilde{R}$ of the harmonic potential is a fitting quantity. Computer simulations show that for N>>1, the value of $\tilde{\rho}_s(0)$ ceases to depend on $N$ and becomes equal to some constant fraction of $\rho_s$. Recalling that in the cases of interest $R_g^2=\frac{1}{2}\tilde{R}b$ we obtain, using the relation (21), that the required relation between $\tilde{\rho}_s(0)$ and $\rho_s$ is achieved if the parameter $\tilde{R}^2$ has the following structure:

$$\tilde{R}^2 = \frac{\beta}{b^2}\left(\frac{N}{\rho_s}\right)^{4/3}, \quad (25)$$

where $\beta$ is a dimensionless numerical coefficient of order 1, possibly reflecting details of the chemical structure of the Kuhn segment. The latter corresponds to the following harmonic potential:

$$U\{\vec{r}_n\} = \frac{3k_B T}{2}\sum_n \frac{\rho_s^{4/3} b^2}{\beta N^{4/3}}\left(\vec{r}_n - \vec{r}_{cm}\right)^2 \quad . \quad (26)$$

Note that this potential is actually very weak, because N>>1. Using relations (22) and (25) we obtain

$$\tilde{\rho}_s(0) = \left(\frac{3}{\pi\sqrt{\beta}}\right)^{3/2} \rho_s . \quad (27)$$

The concentration of the intrinsic ring segments inside the globule becomes comparable to the total concentration of Kuhn segments in the melt $\tilde{\rho}_s \propto \frac{N}{R_g^3} \propto \rho_s$ and does not depend on *N*. For an infinitely large Gaussian ring the concentration of self-segments is much smaller and depends on *N* as $N^{-1/2}$, qualitatively similar behavior was observed in recent computer simulations [4,5]. We can estimate the value of the parameter $\beta$ using the above-mentioned experimental data for melts of ring macromolecules PEO $\tilde{\rho}_s(0) \approx 0.385\rho_s$, and obtain $\beta = \left(\frac{3}{\pi}\right)^2\left[\frac{\rho_s}{\tilde{\rho}_s(0)}\right]^{4/3} \approx 3.25$. This result looks reasonable.

Figure 3 illustrates the dependence of the square of the radius of gyration on the molecular mass of a cyclic macromolecule with potential (26). Instead of the concentration of Kuhn segments $\rho_s$ in the melt, the so-called packing length of the melt $p$ is used hereafter. Both these quantities are related by definition by the following relation [26]:

$$\frac{1}{p} \equiv \rho_s b^2 . \quad (28)$$

Using this relation one finds $\tilde{R}^2 = \beta N^{4/3}\left(\frac{p}{b}\right)^{4/3} b^2$ and $R_g^2 = \frac{\beta^{1/2}}{2} N^{2/3}\left(\frac{p}{b}\right)^{2/3} b^2$. Note also that the characteristic ratio $\frac{p}{b}$ also appears in expression for the number between neighboring entanglements in polymer melts $N_e \approx \left(\frac{21.3p}{b}\right)^2$, what can be easily verified using experimental data published in [26].

.

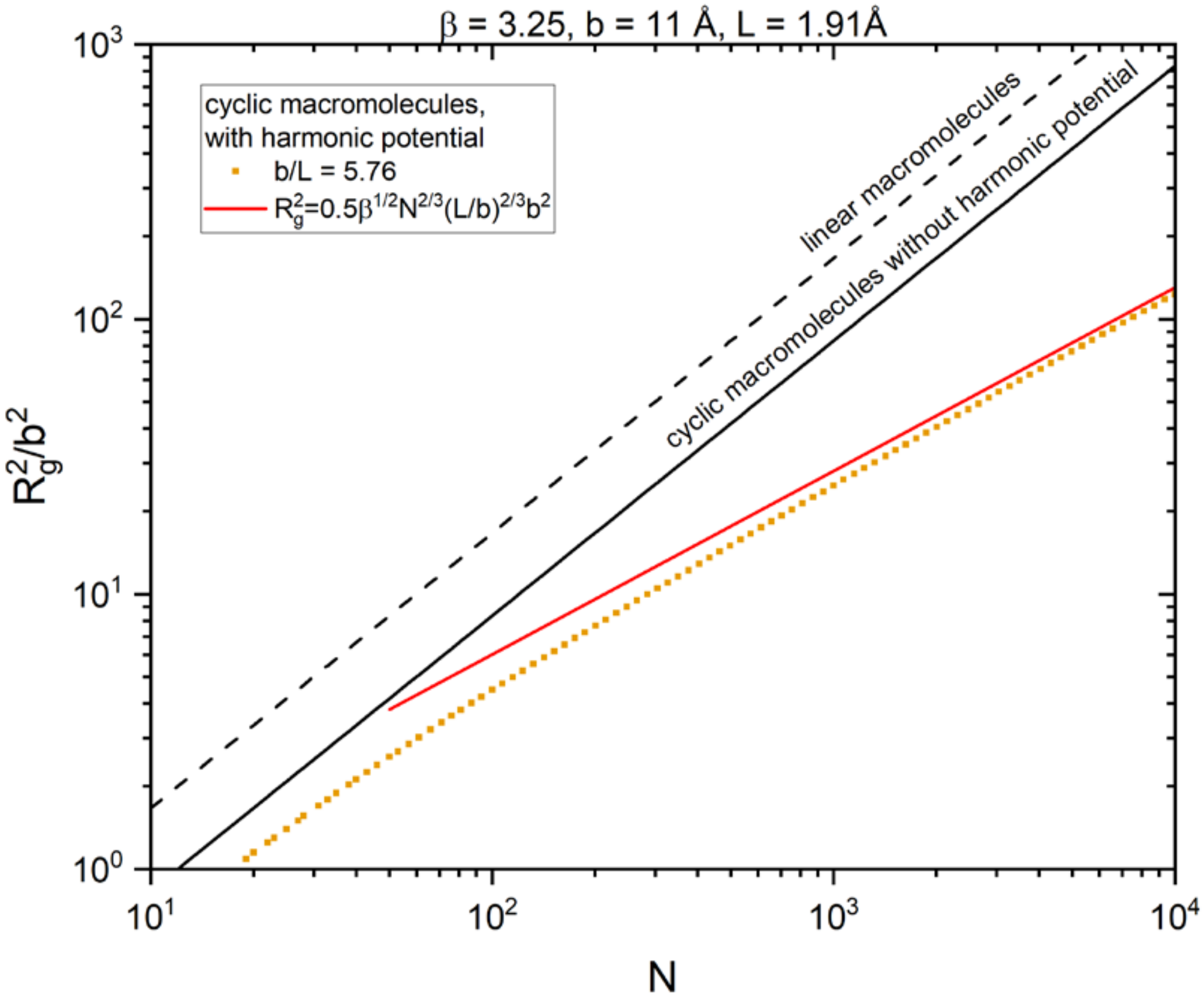

Fig.3. Graphical illustration of the relations (13)-(16). In this case, the harmonic potential parameter $\tilde{R}$ and the number of Kuhn segments N in the cyclic macromolecule are related to each other through the relation (25), $p = L$.

Another feature of the model is related to the fact that the attractive potential (1) does not change the self-diffusion coefficient of the polymer ring, but strongly shortens the terminal relaxation time compared to the Rouse relaxation time, see relation (12):

$$\tau_1 \cong \overline{\tau}_1 = \pi^2 \frac{\tau_s \tilde{R}^2}{b^2} = 4\pi^2 \tau_s \frac{R_g^4}{b^4} << \tau_R{}^2 = \tau_s N^2 . \qquad (29)$$

At times of order $\tau_1$, the mean squared displacements of the ring segments reach saturation, asymptotically approaching $2\left\langle R_g^2 \right\rangle_{eq} = \tilde{R}b$, while the mean squared displacement of the center of mass is even much smaller and only at much larger times $t >> \overline{\tau}_1^* = \dfrac{\overline{\tau}_1}{4}\dfrac{Nb^2}{R_g^2} = 3\overline{\tau}_1 \dfrac{R_{g,ideal}^2}{R_g^2}$ begins to dominate, see Fig.2. For the potential (26) one can find from the above expressions that $\dfrac{\overline{\tau}_1^*}{\tau_1} = \dfrac{1}{2\sqrt{\beta}} N^{1/3} \left(\dfrac{b}{p}\right)^{2/3}$. Effectively, it will lead to the fact that the transition to the normal diffusion regime will be protrated for an extended time interval $\overline{\tau}_1 \le t \le \overline{\tau}_1^*$, during which the time dependence of the mean-squared displacement of the segments will slow down demonstrating a quasi-plateau for large N. Note also that the anomalous behavior on the normal diffusion regime in melts of cyclic macromolecules has been repeatedly noted in computer experiments (see, for example, [4,12]). Probably, though simplified, it is reflected already in the discussed strongly simplified model. A more specific discussion of the latter aspect, however, requires a more realistic description of intramolecular and intermolecular entanglement effects than is provided by the terms associated with the friction forces and the stochastic Langevin force in equation (3). Such studies are planned in future works of the authors.

**References.**

1. Rubinstein, M. Dynamics of Ring Polymers in the Presence of Fixed Obstacles. Phys. Rev. Lett. 1986, 57, 3023–3026, DOI: 10.1103/PhysRevLett.57.3023
2. Kruteva, M.; Allgaier, J.; Richter, D. Topology Matters: Conformation and Microscopic Dynamics of Ring Polymers. *Macromolecules* **2023**, 56, 7203–7229, DOI: 10.1021/acs.macromol.3c00560
3. Ge, T.; Panyukov, S.; Rubinstein, M. Self-Similar Conformations and Dynamics in Entangled Melts and Solutions of Nonconcatenated Ring Polymers. *Macromolecules* **2016**, 49, 708−722, DOI: 10.1021/acs.macromol.5b02319
4. Dell, Z. E.; Schweizer, K. S. Intermolecular Structural Correlations in Model Globular and Unconcatenated Ring Polymer Liquids. *Soft Matter* **2018**, 14, 9132−9142, DOI: 10.1039/C8SM01722K
5. Mei, B.; Dell, Z. E.; Schweizer, K. S. Microscopic Theory of Long-Time Center-of-Mass Self-Diffusion and Anomalous Transport in Ring Polymer Liquids. *Macromolecules* **2020**, 53, 10431−10445, DOI: 10.1021/acs.macromol.0c01737
6. Halverson, J. D.; Lee, W. B.; Grest, G. S.; Grosberg, A. Y.; Kremer, K. Molecular dynamics simulation study of nonconcatenated ring polymers in a melt. I. Statics. *J. Chem. Phys.* **2011**, 134, 204904., DOI: 10.1063/1.3587137

7. Halverson, J. D.; Lee, W. B.; Grest, G. S.; Grosberg, A. Y.; Kremer, K. Molecular dynamics simulation study of nonconcatenated ring polymers in a melt. II. Dynamics. *J. Chem. Phys.* **2011**, 134, 204905. DOI 10.1063/1.3587138

8. Tsalikis, D. G.; Mavrantzas, V. G.; Vlassopoulos, D. Analysis of Slow Modes in Ring Polymers: Threading of Rings Controls Long-Time Relaxation. *ACS Macro Lett.* **2016**, 5, 755−760, DOI: 10.1021/acsmacrolett.6b00259

9. Kruteva, M.; Allgaier, J.; Monkenbusch, M.; Porcar, L.; Richter, D. Self-similar Polymer Ring Conformations Based on Elementary Loops: A Direct Observation by SANS. *ACS Macro Lett.* **2020**, 9, 507–511, DOI: 10.1021/acsmacrolett.0c00190

10. Kruteva, M.; Monkenbusch, M.; Allgaier, J.; Holderer, O.; Pasini, S.; Hoffman, I.; Richter, D. Self-Similar Dynamics of Large Polymer Rings: A Neutron Spin Echo Study. *Phys. Rev. Lett.* **2020**, 125, 238004, DOI: 10.1103/PhysRevLett.125.238004

11. Kruteva, M.; Allgaier, J.; Monkenbusch, M.; Hoffmann, I.; Richter, D. Structure and dynamics of large ring polymers. *J. Rheol.* **2021**, 65, 713–727, DOI: 10.1122/8.0000206

12. Halverson, J. D.; Kremer, K.; Grosberg, A. Y. Comparing the results of lattice and off-lattice simulations for the melt of nonconcatenated rings. *J. Phys. A: Math. Theor.* **2013**, 46, 065002, DOI: 10.1088/1751-8113/46/6/065002

13. Pasquino, R.; Vasilakopoulos, T. C.; Cheol Jeong, Y.; Lee, H.; Rogers, S.; Sakellariou, G.; Allgaier, J.; Takano, A.; Brás, A. R.; Chang, T.; Gooßen, S.; Pyckhout-Hintzen, W.; Wischnewski, A.; Hadjichristidis, N.; Richter, D.; Rubinstein, M.; Vlassopoulos, D. Viscosity of Ring Polymer Melts. *ACS Macro Lett.* **2013**, 2, 874−878, DOI: 10.1021/mz400344e

14. Tu, M. Q.; Davydovich, O.; Mei, B.; Singh, P. K.; Grest, G. S.; Schweizer, K. S.; O'Connor, T. C.; Schroeder, C. M. Unexpected Slow Relaxation Dynamics in Pure Ring Polymers Arise from Intermolecular Interactions. *ACS Polym. Au* **2023,** 3, 307-317, DOI: 10.1021/acspolymersau.2c00069

15. K. L. Ngai, On the Z5.8-dependence of viscosity of polymer rings over the range 15 < Zw ≤ 300 of the entanglement number, J Polym Sci. **2024**;62:174–181, DOI: 10.1002/pol.20230624 .

16. Shota Goto; Kang Kim, Nobuyuki Matubayasi, Effects of chain length on Rouse modes and non-Gaussianity in linear and ring polymer melts, *J. Chem. Phys.* 155, 124901 (**2021**), https://doi.org/10.1063/5.0061281

17. Yamakawa, Hiromi, Modern Theory of Polymer Solutions, Electronic Edition. Harper & Row, **1971**, 434p.

18. M. Doi, S. F. Edwards, The Theory of Polymer Dynamics (Oxford Clarendon Press, **1989**).

19. A. Y. Grosberg, A. R. Khokhlov, Statistical Physics of Macromolecules (AIP Press, **1994**).

20. M. Rubinstein, R. H. Colby, Polymer Physics (Oxford University Press, 2003).

21. P. G. de Gennes, Scaling Concepts in Polymer Physics (Cornell University Press, Ithaka, **1979**).

22. G. R. Strobl, The Physics of Polymers (Springer, **1997**).
23. A. Denissov, M. Kroutieva, N. Fatkullin, R. Kimmich, Segment diffusion and nuclear magnetic resonance spin-lattice relaxation of polymer chains confined in tubes: Analytical treatment and Monte Carlo simulation of the crossover from Rouse to reptation dynamics J. Chem.Phys. **2002**, 116, 5217, https://doi.org/10.1063/1.1451242.
24. Maxim Dolgushev, Margarita Krutyeva, Polymer Chain Confined Into a Harmonic Radial Potential, Macromol. Theory Simul. **2012**, 21, 565–572, DOI: 10.1002/mats.201200045
25. Kevin Lindt, Nail Fatkullin, Carlos Mattea, Jürgen Allgaier, Siegfried Stapf, Margarita Kruteva, Spin Relaxation and Dynamics of Ring Poly (ethylene oxide) in Melts, Macromolecules 57(8), **2024**, DOI:10.1021/acs.macromol.3c02360
26. Fetters, L. J.; Lohse, D. J.; Richter, D.; Witten,T. A.; Zirkel, A., Macromolecules 1994, 27, 4639.